\documentclass[aps,pre,floatfix,superscriptaddress,showpacs,nofootinbib,amsmath,amssymb,reprint]{revtex4-2}
\usepackage[utf8]{inputenc}
\usepackage{float}
\usepackage[caption = false]{subfig}
\usepackage{amsmath}
\usepackage{amssymb}
\usepackage{graphicx}
\usepackage{verbatim}
\usepackage{xcolor}
\usepackage{epstopdf}
\usepackage{bbm}
\usepackage{url}
\usepackage{tikz}
\usepackage{comment}
\usepackage[pdftex]{hyperref}
\usepackage[normalem]{ulem}

\begin{document}

\title{Ambient air plasma acceleration in tightly-focused ultrashort infrared laser beams}

\author{Marianna Lytova}
\email[]{marianna.lytova@inrs.ca}
\affiliation{INRS-EMT, 1650 Boul. Lionel-Boulet, Varennes, Quebec J3X 1P7, Canada}

\author{Fran\c{c}ois Fillion-Gourdeau}
\affiliation{INRS-EMT, 1650 Boul. Lionel-Boulet, Varennes, Quebec J3X 1P7, Canada}

\author{Simon Valli\`eres}
\affiliation{INRS-EMT, 1650 Boul. Lionel-Boulet, Varennes, Quebec J3X 1P7, Canada}

\author{Sylvain Fourmaux}
\affiliation{INRS-EMT, 1650 Boul. Lionel-Boulet, Varennes, Quebec J3X 1P7, Canada}

\author{Fran\c{c}ois L\'egar\'e}
\affiliation{INRS-EMT, 1650 Boul. Lionel-Boulet, Varennes, Quebec J3X 1P7, Canada}

\author{Steve MacLean}
\affiliation{INRS-EMT, 1650 Boul. Lionel-Boulet, Varennes, Quebec J3X 1P7, Canada}
\affiliation{Infinite Potential Laboratories, 485 Wes Graham Way, Waterloo, N2L 6R2, ON, Canada}

\date{\today}
\begin{abstract}

Recent experimental and theoretical results have demonstrated the possibility of accelerating electrons in the MeV range by focusing tightly a few-cycle laser beam in ambient air \cite{vallieres2022}.
Using Particle-In-Cell (PIC) simulations, this configuration is revisited within a more accurate modelling approach to analyze and optimize the mechanism responsible for electron acceleration. In particular, an analytical model for a linearly polarized tightly-focused ultrashort laser field is derived and coupled to a PIC code, allowing us to model the interaction of laser beams reflected by high-numerical aperture mirrors with laser-induced plasmas. A set of 3D PIC simulations is performed where the laser wavelength is varied from 800 nm to 7.0 $\mu$m while the normalized amplitude of the electric field is varied from $a_{0} = 3.6$ to $a_{0} = 7.0$. The preferential forward acceleration of electrons, as well as the analysis of the laser intensity evolution in the plasma and data on electron number density, confirm  that the relativistic ponderomotive force is responsible for the acceleration. We also demonstrate that the electron kinetic energy reaches a maximum of $\approx1.6$ MeV when the central wavelength is of 2.5 $\mu$m.

\end{abstract}

\maketitle
\section{Introduction}

For more than 40 years, the acceleration of plasma electrons in a laser field has attracted the attention of researchers, both because it is driven by fundamental laser-matter interactions and because it may lead to broad practical applications. 
Thus, mechanisms based on the nonlinear ponderomotive force \cite{Tajima} such as laser wakefield acceleration (LWFA) \cite{Lu, Esarey}, direct laser acceleration (DLA) \cite{Gahn,Arefiev_Robinson_Khudik_2015,Hussein_2021,doi:10.1126/sciadv.adk1947} and vacuum laser acceleration (VLA) \cite{Esarey95, Malka, Quesnel,Carbajo, app3010070, 10.1063/1.4738998, PhysRevX.10.041064,powell2024relativistic} were proposed theoretically and implemented experimentally. Owing to the high field strengths that can be reached in recent laser infrastructures, with intensities on the order of 10$^{18}-10^{23}$~W/cm$^2$ \cite{Bahk:04, Yoon:2021}, these techniques can accelerate electrons to relativistic energies on small length scales (few tens of $\mu$m). In the best scenarios, electrons can reach up to the GeV level \cite{Aniculaesei, Babjak}. To push these limits even further and to take advantage of their peculiar field structure, tightly-focused configurations have been investigated to exploit the highest intensity range \cite{PhysRevLett.88.095005,Bochkarev_2007,10.1063/1.3139255,10.1063/1.2830651,10.1063/5.0150260,Zheng:22} 

Tightly-focused ultrashort laser pulses are also needed for the generation of directed ultrashort relativistic electron beams in the MeV range, which are in demand in medicine and other fields of application \cite{Wen19, vallieres2022}. Recently, relativistic electrons with energies up to 1.4 MeV have been observed when a short linearly polarized laser pulse is tightly focused in ambient air by a high numerical aperture parabolic mirror \cite{vallieres2022}. This surprising result was attributed to both the unimpeded propagation of the laser beam, from the mirror to the interaction region, and the ponderomotive acceleration mechanism. The particular geometry of the tightly focused configuration was instrumental to combine these two phenomena. A set of PIC simulations was performed to validate these experimental findings. However, the laser field model used in these simulations was based on the Gaussian beam solution, which is only valid in the paraxial limit when the focusing is not too strong, and included simplifications related to plasma pre-ionization. In this article, the numerical simulations performed in Ref. \cite{vallieres2022} are revisited using a more accurate laser field model that can accommodate for tightly focused configurations and in which the ionization and acceleration of the plasma occurs self-consistently.   

Such accurate numerical models are required to optimize the characteristics of the generated relativistic electron beam, such as the maximum energy of accelerated particles and their angular distribution. The self-consistent system of Vlasov-Maxwell equations is one of the most accurate theoretical models for describing the dynamics of fields and currents in a collisionless plasma \cite{Lifshitz1981}, but its numerical processing entails large computational costs, primarily due to its high dimensionality. Since the 1960s, PIC codes have been commonly used to solve these equations \cite{Dawson} as they are computationally efficient and reasonably accurate when the medium is in the right density range. The cases considered here fall into this category, which is why the PIC code Smilei \cite{Smilei18} is adopted for all the calculations presented in this article. 

Coupling the plasma to intricate tightly focused laser beam models presents the first challenge of our simulations because the laser models usually implemented in PIC codes are variations of Gaussian beams, which are not accurate to represent our configuration. Rather, a full non-perturbative solution has to be embedded in the numerical tools. We argue that this can be performed via the ``initial-value method'', whereby the initial conditions in the Maxwell solver are evaluated from an exact analytical field model with support in a vacuum subregion. The second challenge is the field injection into the gas target. Tightly focused fields have a quasi-spherical wave front. We choose a spherical gas target, which naturally conforms to the wave front and reduces artificial dephasing in the laser propagation, in contrast to more standard cases using plasma slabs. The combined used of these techniques allows us to simulate the acceleration process and to optimize it by varying different parameters, such as the wavelength and intensity of the laser. 

This article is organized as follows. In Section~\ref{vacuum} we obtain an exact analytical solution for a tightly-focused linearly polarized (LP) pulse, incorporate it into the PIC code's Maxwell equations solver, and discuss the size requirements for the simulation box and the gas target for further numerical modeling in plasma. In Section~\ref{3D}, we present simulation results of electron beams accelerated by lasers with different central wavelengths and intensities, and discuss the possible acceleration mechanism. General conclusions are drawn in the final section \ref{concl}.

\section{Tightly-focused field in vacuum and their embedding in PIC-codes}\label{vacuum}

\begin{figure}[h]
	\centering{
		\includegraphics[scale=0.9,trim={10 20 10 20},clip]{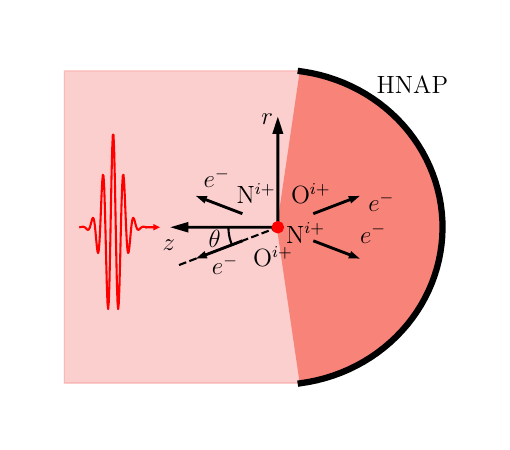}}
	\caption{Sketch of an experimental setup for laser acceleration of electrons ($e^{-}$) in air. The ultra-short infrared (IR) laser pulse (shown as the red envelope on the left) is tightly focused using a High Numerical Aperture Parabola (HNAP). Under the action of this pulse, ionization of air molecules occurs (N$^{i+}$, O$^{i+}$) near the focus (shown as a red dot), and the released electrons can be accelerated.}  \label{hnap}
\end{figure}

The physical system considered in this article is the same as the one studied in Ref. \cite{vallieres2022}. As depicted in Fig.~\ref{hnap}, it consists of a short laser pulse tightly-focused by a parabolic mirror in ambient air. As the pulse propagates from the reflector to the focus, the intensity increases gradually up to a point where ionization begins and electrons are released to form a plasma. Our approach to model this physical system is now described. 

\subsection{Exact solution for ultrashort LP tightly-focused pulse in vacuum}\label{exact_LP}

The experimental configuration (Fig.~\ref{hnap}) simulated in this article has a high numerical apperture (NA) mirror, resulting in a tightly focused beam with a large divergence angle. In this regime, the standard approach based on the paraxial approximation is not applicable and therefore, other methods need to developed. In the past, numerical \cite{Jeong:15, Dumont_2017, Vallieres:23, Majorosi:23} and analytical \cite{April_2010, Salamin:06, Salamin15, Zheng_22} solutions have been considered to model ultrashort tightly-focused pulses in vacuum. However, the numerical techniques usually have an important computational time overhead while the known analytical approaches do not satisfy the requirements of LP and very high-NA. For this reason, we derive a closed-form solutions of Maxwell's equations for a tightly-focused linearly polarized beam based on April's technique \cite{April_2010}. 
Within this approach, the EM field components are expressed in terms of the electric ${\bf \Pi}_{\mathrm{e}}$ and magnetic ${\bf \Pi}_{\mathrm{b}}$ Hertz potentials
\begin{align}
	{\bf E}&=\nabla\times\nabla\times{\bf \Pi}_{\mathrm{e}} -\partial_t\nabla\times{\bf \Pi}_{\mathrm{b}},\\
	{\bf B}&=\nabla\times\nabla\times{\bf \Pi}_{\mathrm{b}}+c^{-2}\partial_t\nabla\times{\bf \Pi}_{\mathrm{e}}.
\end{align}
It is assumed that in vacuum both potentials satisfy the wave equation with a phase velocity $c$ equal to the speed of light. In the case of a pulse polarized in the $x$-direction and propagating in the $z$-direction, they can be written as
\begin{align}
	{\bf \Pi}_{\mathrm{e}} &= {\bf e}_x \Psi, \\
	{\bf \Pi}_{\mathrm{b}} &= {\bf e}_y c^{-1} \Psi,
\end{align}
where $\Psi(x,y,z,t)$ is a scalar function to be determined later and ${\bf e}_{x,y}$ are unit vectors in direction $x$ and $y$, respectively. Thus, the components are given by \cite{April_2010}
\begin{eqnarray}\label{components}	
\nonumber E_x&=&\partial^2_{xx}\Psi-c^{-2}\partial^2_{tt}\Psi+c^{-1}\partial^2_{tz}\Psi,\\
\nonumber E_y &=& \partial^2_{xy}\Psi, \\
\nonumber E_z&=&\partial^2_{zx}\Psi-c^{-1}\partial^2_{tx}\Psi,\\ 
 cB_x &=& \partial^2_{xy}\Psi, \\
\nonumber cB_y&=&\partial^2_{yy}\Psi-c^{-2}\partial^2_{tt}\Psi+c^{-1}\partial^2_{tz}\Psi, \\
\nonumber cB_z&=&\partial^2_{zy}\Psi-c^{-1}\partial^2_{ty}\Psi.
\end{eqnarray}
Further, for an \textit{isodiffracting} pulse, i.e. when all frequency components have the same wavefront radius of curvature \cite{Melamed:98}, the required function $\Psi$ has the form \cite{April_2010}
\begin{equation}\label{phasor}
	\Psi(x,y,z,t) = \dfrac{\Psi_0}{\widetilde{R}}[f(\tilde{t}_+)-f(\tilde{t}_-)],
\end{equation}
where $f(t)$ is related to the spectral content of the pulse.
Thus, the third ingredient of April's method is a Poisson-like spectrum for ultrashort pulses, so that only positive frequencies are present. 
In this case, for a frequency $\omega_0 = k_0c = 2\pi c/\lambda_0$, where $\lambda_0$ is the central wavelength, and constant phase $\phi_0$, the function from \eqref{phasor} has the form \cite{April_2010}
\begin{equation}\label{poisson}
	f(t)= e^{i\phi_0}\Big(1-\dfrac{i\omega_0t}{s}\Big)^{-(s+1)},
\end{equation}
whose arguments
\begin{equation}
	\tilde{t}_\pm = t \pm \widetilde{R}/c+ib/c,
\end{equation}
are given in terms of the complex spherical radius
\begin{align}\label{til_R}	
	\widetilde{R}=\sqrt{r^2+(z+ib)^2},
\end{align}
where $r^2 = x^2+y^2$. 
In expressions \eqref{phasor}-\eqref{til_R}, the real parameters $\Psi_0$, $b$ and $s$ are associated to the pulse amplitude, focus spot size (so-called confocal parameter), and envelope duration, respectively (same as radially polarized pulses described in Ref. \cite{Marceau:12}). There is no need to apply an envelope separately as it is already built into the analytical solution. Moreover, the maximum intensity at the focus $x=y=z=0$ is reached when $t=0$.

Using the relationships between the confocal parameter $b$ and the beam divergence angle $\delta$: $k_0b~=~2/\sin\delta\tan\delta$, and between the Numerical Aperture (NA), and $\delta$: NA$=\sin\delta$, we can express $b$ as a function of NA:
\begin{align}\label{b_NA}
	b = \frac{2\sqrt{1-\text{NA}^2}}{k_0\text{NA}^2}.
\end{align}
In its turn, parameter $s$ is responsible for the number of cycles in the envelope or its duration. Thus, it is related to the full width at half maximum for intensity by \cite{Caron}
\begin{align}\label{tau_w0}
	\tau_{\text{FWHM}}=\sqrt{2}s\sqrt{2^{2/(s+1)}-1}/\omega_0.
\end{align}

Substituting \eqref{phasor} to \eqref{components}, we obtain the field components
\begin{equation}\label{ExEyEzBxByBz}
	\begin{array}{rclcrcl}		
	E_x &=&\dfrac{\Psi_0}{b^3}\Big(\dfrac{x^2}{r^2}\mathcal{K}_1-\mathcal{K}_2\Big),&& cB_x &=&\dfrac{\Psi_0}{b^3}\dfrac{xy}{r^2}\mathcal{K}_1,\\
	 \\
	E_y &=& \dfrac{\Psi_0}{b^3}\dfrac{xy}{r^2}\mathcal{K}_1,&& cB_y &=& \dfrac{\Psi_0}{b^3}\Big(\dfrac{y^2}{r^2}\mathcal{K}_1-\mathcal{K}_2\Big),\\
	\\ E_z &=&\dfrac{\Psi_0}{b^3}\dfrac{x}{r}\mathcal{K}_3,&&	cB_z &=& \dfrac{\Psi_0}{b^3}\dfrac{y}{r}\mathcal{K}_3,
	\end{array}	
\end{equation}
written in terms of
\begin{align*}
	\mathcal{K}_1=& \sin^2\tilde{\theta}\Big(\dfrac{3\mathcal{G}^{(0)}_-}{\widetilde{\mathcal{R}}^3}-\dfrac{3\mathcal{G}^{(1)}_+}{\widetilde{\mathcal{R}^2}}+ \dfrac{\mathcal{G}^{(2)}_-}{\widetilde{\mathcal{R}}} \Big),\\
	\\
	\mathcal{K}_2=&\dfrac{\mathcal{G}^{(0)}_-}{\widetilde{\mathcal{R}}^3}+\dfrac{\mathcal{G}^{(1)}_-\cos\tilde{\theta}}{\widetilde{\mathcal{R}^2}} -\dfrac{\mathcal{G}^{(1)}_+}{\widetilde{\mathcal{R}^2}}+\frac{\mathcal{G}^{(2)}_-}{\widetilde{\mathcal{R}}}-\mathcal{G}^{(2)}_+\dfrac{\cos\tilde{\theta}}{\widetilde{\mathcal{R}}},\\
	\\
	\mathcal{K}_3=&\sin\tilde{\theta}\Big(\dfrac{3\mathcal{G}^{(0)}_-\cos\tilde{\theta}}{\widetilde{\mathcal{R}}^3}+\dfrac{\mathcal{G}^{(1)}_-}{\widetilde{\mathcal{R}^2}}-\dfrac{3\mathcal{G}^{(1)}_+\cos\tilde{\theta}}{\widetilde{\mathcal{R}^2}}+\\
	&\qquad\qquad\qquad\frac{\mathcal{G}^{(2)}_-\cos\tilde{\theta}}{\widetilde{\mathcal{R}}}- \frac{\mathcal{G}^{(2)}_+}{\widetilde{\mathcal{R}}} \Big).
\end{align*}
Here, the functions
\begin{equation}
	\mathcal{G}_\pm^{(n)}=g^{(n)}(\tilde{\tau}_+)\pm g^{(n)}(\tilde{\tau}_-),\quad n=0,1,2
\end{equation} 
are expressed through the following function ($n=0$) or derivatives ($n=1,2$)
\begin{equation}\label{poisson_dstr}
	g^{(n)}(\tau)=e^{i\phi_0}\dfrac{\Gamma(s+n+1)}{\Gamma(s+1)}\Big(\dfrac{ik_0b}{s}\Big)^n\Big(1-\dfrac{i\tau}{s}\Big)^{-(s+n+1)},
\end{equation}
with arguments
\begin{equation}
	\tau_\pm = \tau + k_0b(\pm\widetilde{\mathcal{R}}+i),
\end{equation}
where $\tau = \omega_0t$, $\widetilde{\mathcal{R}} = \sqrt{\rho^2+(\zeta + i)^2}$, $\rho= r/b$, $\zeta = z/b$. Also in the expressions for $\mathcal{K}_{1,2,3}$ we use the notation: $\sin\tilde{\theta}=\dfrac{\rho}{\widetilde{\mathcal{R}}}$, $\cos\tilde{\theta} = \dfrac{\zeta + i}{\widetilde{\mathcal{R}}}$. Thus, the solution \eqref{ExEyEzBxByBz} is only approximately LP, being purely LP only on the optical axis, since we are dealing with a finite beam \cite{John_Lekner_2003}.

To fix $\Psi_0$, we use \eqref{ExEyEzBxByBz} and the fact that at the focal spot $x=y=z=0$ at time $t=0$, the peak intensity in vacuum is $I_0 = c\varepsilon_0(\mathrm{Re}\{E_{x}^{\mathrm{foc}}\})^2/2=\Psi_0^2(\mathrm{Re}\{\mathcal{K}_{2}^{\mathrm{foc}}\})^2/2\mu_0c b^6$,
where for $\mathcal{K}_{2}^{\mathrm{foc}}$, we can obtain the analytical expression
\begin{align}\label{K2_max}
	\nonumber\mathcal{K}_{2}^{\mathrm{foc}}&=ie^{i\phi_0}\Big[(1+2k_0b/s)^{-(s+1)}-1+\\
	&2k_0b(s+1)/s-2k_0^2b^2(s+2)(s+1)/s^2 \Big].
\end{align}
Hence, in the end, the amplitude is given by
\begin{equation}\label{psi0_int}
	\Psi_0 = \frac{\sqrt{2\mu_0cb^6I_0}}{\mathrm{Re}\{\mathcal{K}_{2}^{\mathrm{foc}}\}}.
\end{equation}

\subsection{Modeling approaches}\label{models}

To model the acceleration of particles in the desired field configuration given in the previous section, we first need to incorporate this solution into the Maxwell's equation solver. This can be performed in a number of ways. For example, in \cite{PhysRevE.104.015203}, a short tightly-focused laser pulse is evaluated numerically from a field calculator based on Stratton-Chu integrals. This pulse is then used as an initial condition in PIC simulations, positioned beside the plasma region and thus, requiring larger simulation boxes. In Ref. \cite{THIELE20161110}, a different approach is presented where the field is prescribed on a plane in the simulation box and back-propagated to the boundary. In Ref. \cite{Wen19}, an approximate stationary analytic solution for the non-paraxial case multiplied by some specific time-envelope was implemented in the scattered field formalism (SFF), where the total field is divided into in-vacuum and scattered parts \cite{Kunz}. Recently a similar approach, referred to as the analytic pulse technique, was justified in \cite{weichman2023analytic} for implementation with the PIC algorithm. The authors of the latter paper analyze the limitations caused by numerical and physical dispersion and provide simulation examples for relatively long 100 fs pulses containing a large number of cycles. Finally, the SFF has also been used to study particle acceleration in low density gas using longitudinal fields,  \cite{PhysRevLett.111.224801,Marceau_2015}. Therefore, there are several options for creating a laser in PIC codes:  
\begin{enumerate}
	\item[(i)] Specify time-dependent boundary conditions for the magnetic field components on the faces (in general, on all 6 of them in 3-D) of the simulation box. Note that the region of the medium that can interact with the electromagnetic (EM) field must be at least several wavelengths away from these faces.
	\item[(ii)] Specify an initial condition representing the laser beam at time $t_{\mathrm{initial}}$ for vacuum points in a subregion of the simulation box, i.e. before entering the plasma region, and let it evolve through the domain.  
	\item[(iii)]Implement the scattered field formalism \cite{Kunz} by specifying all components of a known \textit{incident} vacuum field in space and time and using the Maxwell solver to calculate the scattered field only.
\end{enumerate}
In principle, each of the approaches (i)-(iii) is applicable for our purposes. We now look at the pros and cons of each method to present and justify our choice. 

As usual with 3-D PIC modeling, we are faced with limited computing resources to model the problem in the medium. It would be physically reasonable, but computationally expensive, if possible at all, to start in neutral gas far from the focus to model how the pulse propagates, intensifies as it approaches the focus, and simultaneously decays ionizing the gas and interacting with plasma waves.
Thus, the correct choice of simulation model plays a crucial role. Long-term numerical propagation can lead to the accumulation of errors, especially if we consider the laser pulse arising from the faces of the simulation box (method (i)), or as an external field specified at $t_\text{initial} $ (method (ii)). At first glance, the SFF method, mentioned as (iii), is free from this drawback, since it uses an analytical solution for the incident field everywhere and for any time $t$. However, in practice, calculating this solution using formulas \eqref{ExEyEzBxByBz} and sending them to the PIC-code control file at each point in time for each point in space generates large computational overhead. Moreover, in our tests we saw instabilities in the scattered field solutions, which, although can be eliminated by reducing the cell sizes, lead to increased simulation time and resources. As for setting the laser on the boundaries, method (i), this approach also requires field calculations for the 6 box faces at each time step, which turns out to be more expensive than the optimized Maxwell solvers provided by PIC codes. Thus, we settled on option (ii), calculating the initial field only once and letting it evolve according to Maxwell's equations. The accuracy of this method is tested in vacuum in Section \ref{sec:laser_vac}.

\subsection{Selecting laser beam and target parameters}\label{las_param}

\subsubsection{Laser parameters}

The goal of this final section on tightly-focused solutions in vacuum is to set the laser pulse and gas target parameters for subsequent simulations in section~\ref{3D}, where such IR pulses with different central wavelengths interact with air plasma. Beforehand, it is convenient to normalize all variables to units typical in plasma physics, see Table~\ref{tab:units}. Then, to compare similar laser conditions for all $\lambda_0$, we fix the number of cycles to the same value such that $\tau_{\text{FWHM}}\cdot\omega_0=const$. To solve equation \eqref{tau_w0} for $s$, we take from the experiment \cite{vallieres2022} $\lambda_0 = 1.8$ $\mu$m and $\tau_{\text{FWHM}} = 12$~ fs, resulting in $s\approx57.18$. We keep this value constant for all tests in this article, only adjusting $\tau_{\text{FWHM}}$ to match $\lambda_0$.
For example, when $\lambda_0 = 0.8$~$\mu$m and $2.5$ $\mu$m, the corresponding pulse duration is $\tau_{\text{FWHM}} = 5.3$~fs and $16.7$~fs, respectively. Based on the data for the HNAP \cite{Vallieres:23} used in the experiment \cite{vallieres2022}, in all our simulations we assume NA=0.95, hence the confocal parameter is $b\approx0.11\lambda_0$, confirming the non-paraxial regime. In the simulations we also set $\phi_0=\pi/2$ so that the maximum occurs at the focus at time $t=0$, see \eqref{K2_max}.

\begin{table}[b]
	\caption{Normalizing plasma units \label{tab:units}}	
\begin{ruledtabular}
	\begin{tabular}{ l l } 
		Reference quantity & Expression (SI) \\
		\hline 			
		Time ($T_{r}$) & $1/\omega_0$ \\
		Electric fields ($E_r$) & $m_ec/eT_r$ \\  			
		Critical density ($n_c$) & $\varepsilon_0 m_e/e^2T_r^2$		
	\end{tabular}		
\end{ruledtabular}
\end{table}
The maximum kinetic energy of electrons accelerated in the ponderomotive potential is estimated as
\begin{align}\label{max_kin}	\mathcal{E}^{\mathrm{max}}_{K}=m_ec^2(\sqrt{a_0^2+1}-1),
\end{align}
where the normalized amplitude of the electric field vector $a_0= |{\bf E}_0|/E_r=\sqrt{I_0\lambda_0^2(e^2/2\pi^2\varepsilon_0m_e^2c^5)}=\sqrt{I_0(\text{W/cm}^2)\lambda_0(\mu\text{m})^2/1.37\cdot10^{18}}$. This means that to compare the efficiency of electron acceleration by lasers of different wavelengths, it is reasonable to keep the same value $a_0$.  Thus, using \eqref{psi0_int} it is convenient to obtain an expression for the amplitude $\Psi_0$ depending on $a_0$:
\begin{equation}
	\Psi_0 = \frac{a_0k_0b^3m_ec^2}{\mathrm{Re}\{\mathcal{K}_{2}^{\mathrm{foc}}\}e}.
\end{equation}

\begin{figure*}
	\centering{
		\includegraphics[scale=1,trim={0 0 0 0},clip]{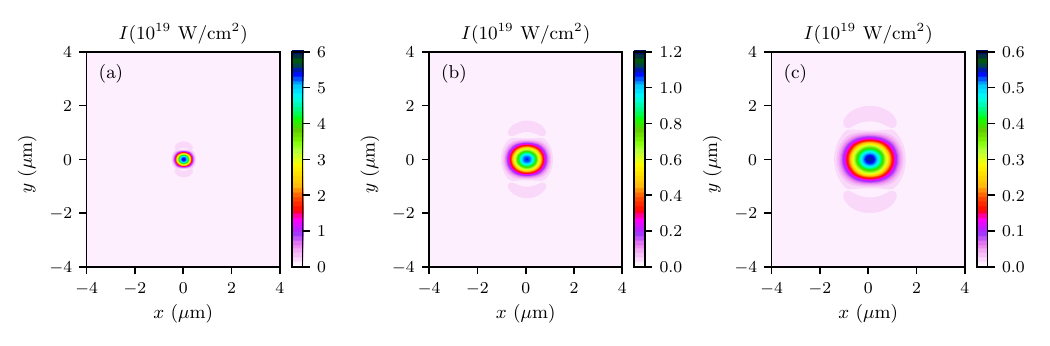}}
	\caption{Analytically calculated intensities $I(x,y)$ in the beam focus plane $z=0$ at $t=0$ for $a_0=5$ and $\lambda_0$ (a) 0.8~$\mu$m ($I_0\approx5.34\cdot10^{19}$ W/cm$^2$), (b) 1.8~$\mu$m ($I_0\approx1.05\cdot10^{19}$ W/cm$^2$), (c) 2.5~$\mu$m ($I_0\approx0.54\cdot10^{19}$ W/cm$^2$). }  \label{focus_xy}
\end{figure*}

\begin{figure}[t]
	\centering{
		\includegraphics[scale=1,trim={0 5 0 5},clip]{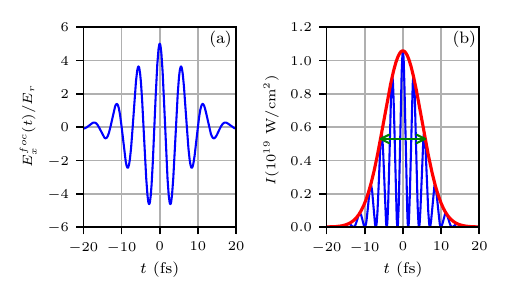}}
	\caption{Analytical solution for LP tightly focused pulse in vacuum: (a) Electric field strength in normalized units at the beam focus over time; (b) The corresponding intensity (in blue) and its envelope (in red); the horizontal green bar indicates the full width at half maximum of 12 fs. Parameters: $s=57.18$, NA=0.95, $a_0=5$, $\lambda_0=1.8$ $\mu$m.
	}  \label{hilb_env}
\end{figure}

On the other hand, although for given NA and $s$ the normalized field amplitudes do not depend on $\lambda_0$, the peak intensities in W/cm$^2$ and the focal spot sizes in $\mu$m differ for various wavelengths, see Fig.~\ref{focus_xy}(a)-(c). For insert (b), the temporal envelopes of the field and intensity at the focus of the laser beam are shown in Fig.~\ref{hilb_env}(a),(b). Thus, at the same $a_0$, in the air target, we can expect a higher level of near-focus ionization in the case of a pulse with a lower $\lambda_0$.


\subsubsection{Target parameters}

\begin{figure}[b]
	\centering{
		\includegraphics[scale=1,trim={0 12 0 20},clip]{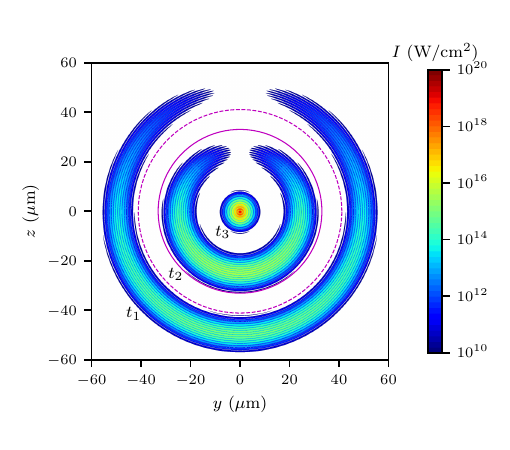}}
	\caption{The intensity $I(0,y,z,t)$ in vacuum calculated from the expressions \eqref{ExEyEzBxByBz} is presented for the parameters $a_0=5$ and $\lambda_0 = 1.8$ $\mu$m. Solutions for three time points are shown: $t_1=-165$ fs, $t_2=-82.5$ fs and $t_3=0$ (focusing time). Here the magenta circles show the  position of the gas target with number density $n_{\mathrm{air}}(x,y,z)$, such that the solid line corresponds to $0.95n_\mathrm{max}$ level, while the dashed line marks the $0.5n_\mathrm{max}$ level.}  \label{pulse_t0}
\end{figure}

The shape and size of the gas target should be selected based on the geometry of the laser field in vacuum and its intensity at the boundary of this target. Thus, from Fig.~\ref{pulse_t0} we conclude that the spherical symmetry of the target ensures that every part of the EM wave travel the same distance in the gas, thereby avoiding artificial losses of intensity at the focus, as would be in the case of a boxy target. 

The question remains how large this sphere should be? In Figure~\ref{pulse_t0}, the maximum intensity at $t_{\mathrm{initial}}=t_1$  is $I^\mathrm{max}(t_1)\approx1.37\times10^{15}$ W/cm$^2$, so we can expect ionization of air molecules up to N$^3+$ and O$^3+$ \cite{vallieres2022}.
To start the simulation where ionization is just beginning, when the pulse has a radius $R_{\mathrm{ion}}$, the laser should have a maximum intensity of $I_\mathrm{ion}\approx2\times10^{14}$ W/cm$^2$. To estimate the value of $R_{\mathrm{ion}}$, we notice  that on the propagation axis, i.e. when $x=y=0$, and in the limit where $z\gg b$, Eq. \eqref{ExEyEzBxByBz} becomes
\begin{align}
    (\mathrm{Re}\{E_x(z,\varphi)\})^2\approx\frac{\mathcal{E}_0^2}{z^2}\mathcal{P}(\varphi,s),
\end{align}
where the factor
\begin{align}
	\mathcal{E}_0 &= \frac{2a_0k_0^3b^3m_ec^2}{\mathrm{Re}\{\mathcal{K}_2^\mathrm{foc}\}e}\frac{(s+2)(s+1)}{s^2}, 
\end{align}
does not depend on $\lambda_0$ but only NA, $s$ and $a_0$, whereas the function
\begin{align}
    \mathcal{P}(\varphi,s) = \sin^2\Big[(s+3)\arctan\Big(\frac{\varphi}{s}\Big)\Big]/\Big(1+\frac{\varphi^2}{s^2}\Big)^{s+3}
\end{align}
depends on the phase in the far field $\varphi=\tilde{\tau}_-(0,0,z,t)=\omega_0t-k_0z$. In the range $\varphi\in[0,2\pi]$, the maximum value $\mathcal{P}^{\mathrm{max}}\approx0.96$, that results in
\begin{align}
	R_\mathrm{ion}=\sqrt{c\varepsilon_0\cdot0.96/2 I_\mathrm{ion} }\cdot\mathcal{E}_0,
\end{align}
which yields $R_\mathrm{ion} \approx128\mu $m.

Using the B-integral, we can measure the influence of nonlinear effects on the propagation of the laser pulse from the mirror to this distance
\begin{align}
	B = \frac{2\pi}{\lambda_0}\int^{R_\mathrm{ion}}_f n_2I(z)dz,
\end{align}
where $n_2$ is the non-linear refractive index, $f=6.35$ mm is the focal length of HNAP.  We estimate the intensity on the propagation axis as $I(z)=c\varepsilon_0\mathcal{E}_0^2\langle\mathcal{P}(s)\rangle_{\varphi\in[0, 2\pi]}/z^2$, with average $\langle\mathcal{P}(s)\rangle_{\varphi\in[0, 2\pi]}\approx0.39$. To evaluate this integral we use $n_2=7.99\cdot10^{-8}$ cm$^2$/TW, assuming only an electronic response with no rotational contribution at such a short pulse duration \cite{Zahedpour:15}, which results in a small value $B=5.7\cdot(1.8/\lambda_0[\mu\mathrm{m}]) \;\mathrm{mrad} \ll 2\pi/10$ rad for all $\lambda_0$, in agreement with the estimations of Ref. \cite{vallieres2022}. Therefore, nonlinear effects outside the simulation box can be neglected for all cases considered in this article, even if the propagation is in ambient air.

Of course, using the target size $R_{\mathrm{target}}=R_\mathrm{ion}$ would significantly increase the computation time and resources for 3D modeling compared to that shown in Fig.~\ref{pulse_t0} where $R_{\mathrm{target}} \approx 40 \; \mu\text{m}$, and even more so in plasma. Fortunately, ionized electrons far from the focal spot do not contribute to the flux of energetic particles because in this region, the field strength and its gradient are insufficient to cause noticeable acceleration. We also see that this local ionization does not increase the number of electrons in the immediate vicinity of the focus, where they could be accelerated later by the focused laser field. Thus, it is reasonable to start the simulation at time $t_1$, in the position shown in the Fig.~\ref{pulse_t0}. Furthermore, to minimize the reflection of the EM wave on the target boundary and permit the beam injection inside the gas, we use a spherical target with a smoothly increasing density at the boundary (super-Gaussian distribution of power $p$):
\begin{align}\label{superG}
	n_\mathrm{gas}(x,y,z)= n_{\mathrm{gas}0}\exp\Big\{-\Big(\dfrac{x^2+y^2+z^2}{2\sigma^2}\Big)^p\Big\},
\end{align}
where in our simulations $p=6$ and the standard deviation $\sigma = 30[\mu\text{m}]~\cdot~\lambda_0[\mu\text{m}]/1.8\mu\text{m}$. To include such a target, along with the initial solution for a laser pulse in a vacuum, we need to use a simulation box of size at least $64\lambda_0\times64\lambda_0\times64\lambda_0$ for all $\lambda_0$. Further empirical tests in plasma with calculations at various target sizes, starting from $30\lambda_0\times30\lambda_0\times30\lambda_0$, showed satisfactory convergence of results when approaching the selected simulation window size, most probably because the electron density near this boundary remains relatively low.

\subsubsection{Laser field in vacuum: numerical propagation}
\label{sec:laser_vac}
Before proceeding with gas/plasma modeling, we need to compare the result of numerical pulse propagation in vacuum with the exact analytical solution. Using ctypes, a well-known Python library, we can implement any desired field configuration as optimized C-functions, which is faster than directly including expressions~\eqref{ExEyEzBxByBz} in the Python Smilei input file. Technically, at time $t_\mathrm{initial}$ we initialize the field components using the $\mathtt{ExternalField}$ blocks of the Smilei PIC code \cite{Smilei18} and then propagate them with the 4th order solver M4\cite{LU2020109388}, since the accuracy of the 2nd order Yee solver typically used for Maxwell's equations turned out to be insufficient in case of tightly-focused field. For the numerical simulations, we use the same box size $64\lambda_0\times64\lambda_0\times64\lambda_0$ - as was done to present the analytical solution in Fig.~\ref{pulse_t0}. The spatial steps are set identical for all directions: $\Delta x= \Delta y = \Delta z = \lambda_0/32$, which, taking into account the Courant–Friedrichs–Lewy stability condition \cite{Strik, gross} leads to $ \Delta t = 0.99\lambda_0/32\sqrt{3}c$.

\begin{figure}[b]
	\centering{
		\includegraphics[scale=1,trim={0 12 0 20},clip]{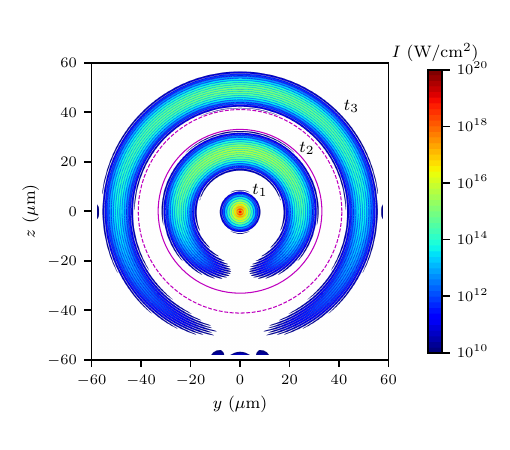}}
	\caption{The intensity $I(0,y,z,t)$ of a LP tightly-focused laser pulse as result of numerical propagation in vacuum for the parameters $a_0=5$ and  $\lambda_0 = 1.8$ $\mu$m. Solutions for three time points are shown: $t_1=0$ (focusing time), $t_2=82.5$ fs and $t_3=165$ fs (symmetric to initial). The magenta circles show the proposed position of the gas target as in Fig.~\ref{pulse_t0}.}  \label{num_prop}
\end{figure}

We start with the field intensity configuration shown on Fig.~\ref{pulse_t0} at $t_{\mathrm{initial}}=-165$ fs, and in Fig.~\ref{num_prop} we demonstrate the numerical propagation results at times $t_1=0$, $t_2=82.5$ fs and $t_3=165$ fs. 
As can be seen, during the process of numerical propagation in vacuum, only a few spurious fields with an intensity slightly higher than 10$^{10}$ W/cm$^2$ appeared near the bottom, right and left boundaries of the computational domain, and artificial diffusion is quite small.  
Thus, all values of maximum intensities obtained for the points of time shown in Fig.~\ref{num_prop}: $I^{\mathrm{max}}(t_1)=I_0=1.054\times10^{19}$~W/cm$^2$, $I^{\mathrm{max}}(t_2)=5.547\times10^{16}$~W/cm$^2$ and $I^{\mathrm{max}}(t_3)=1.366\times10^{15}$~W/cm$^2$ differ from the corresponding analytical values by no more than $0.4\%$, which seems to us accurate enough to move on to plasma calculations.

At the end of this section, we note that for $\lambda_0=1.8$~$\mu$m and $a_0=5$, there is a good agreement between our analytical solutions and the numerical results obtained using the in-house code \textit{Strattocalculator}, based on Stratton-Chu integrals \cite{Dumont_2017}. This applies both to the shape of the solution in the form of a spherical wave propagating towards the focus, and to the size of the focal spot: $w_\text{FWHM}\approx$1 $\mu$m \cite{vallieres2022}, see Fig.~\ref {focus_xy} (b).
This is why we believe that for our numerical simulations, we can use the analytical formulas \eqref{ExEyEzBxByBz} as a computationally cheaper surrogate for the more accurate Stratton-Chu integral-based solutions.

\section{Modeling of electrons acceleration in tightly-focused LP fields}\label{3D}

In this section, the simulation results for electron acceleration from the tightly-focused laser field in ambient air are presented. This is preceded by further details on the setup.

\subsection{Setup of the numerical experiments}\label{setup}
For initial EM fields, the geometry of the numerical experiment corresponds to that shown in Fig.~\ref{pulse_t0} at $t_1$ (for different $\lambda_0$, both spatial and time scales should be multiplied by factor $\lambda_0[\mu\mathrm{m}]/1.8$). The same values for the simulation box, cell size and time step are used as for the propagation in vacuum presented at the end of section ~\ref{las_param}. In all tests we keep the parameters $s=57.18$ and NA=0.95 unchanged, as described in section~\ref{las_param}. Two sets of parameter sweep are performed:
\begin{enumerate}
    \item For a fixed $a_0=5$,  we vary the central wavelength $\lambda_0 = \{0.8, 1.8, 2.5, 3.5, 4.5, 5.5 , 7.0\}$ $\mu$m (and, accordingly, the pulse duration).
    \item For a fixed $\lambda_0=2.5$ $\mu$m, we change the parameter $a_0 = \{3.6, 5.0, 7.0\}$.
\end{enumerate}

We model the spherical air target as consisting of $79\%$ nitrogen and $21\%$ oxygen, with the number density of air molecules estimated as $n_{\mathrm{air}0}=2.687\times10^{25}$ m$^{-3}$ under standard conditions. Thus, assuming pre-dissociation of N$_2$ and O$_2$ molecules, for our gas target we use distributions \eqref{superG} with $n_{\mathrm{N}0}=0.79\cdot2\cdot n_{\mathrm{air}0}$, $n_{\mathrm{O}0}=0.21\cdot2\cdot n_{\mathrm{air}0}$ and $n_{\mathrm{e}0}=0$, so the air is initially neutral. For each type of molecule/ion, we take 1 superparticle per cell between spheres with radii $r_0$ and $r_0/3$ and 2 superparticles within a sphere with radius $r_0/3$, where $r_0$ is chosen such that $n_{\mathrm{gas}}(r_0)=0.001n_{\mathrm{gas}0}$ in \eqref{superG}. In total, the number of N and O superparticles is slightly more than $6\cdot10^{9}$.

When propagating a pulse in a gas, for all the species of ions/molecules, we take into account tunneling ionization by a laser field using the appropriate Smilei module, based on the ADK semi-classical formalism \cite{10.1063/1.3559494}. Thus, depending on the local degree of ionization, the actual number of electron superparticles in a cell varies from 0 to 30, and their total number can even exceed 10$^{10}$. The electron superparticles generated in this way are pushed using the scheme proposed by Higuera and Cary \cite{Higuera_2017} and implemented in the Smilei code.
Unlike electrons, we consider air ions to be at rest due to their large masses. This assumption saves computational time and does not have a significant effect on the electron acceleration, most probably because the plasma frequency for ionic species is much lower than for electrons: $\omega_\text{p,e} \gg \omega_\text{p,N}, \omega_\text{p,O}$. Depending on the total number of electron superparticles, i.e. the level of ionization, the simulations presented further in Section~\ref{3D} run for 6-14 hours on 5120 cores.

Using diagnostic tools provided by Smilei, we measure the following data at several pre-selected points in time
\begin{enumerate}
	\item Energy-angular spectra of accelerated electrons in the pulse polarization plane $(x,z)$ and in the transverse to polarization plane $(y,z)$. 
	
	\item Spatial distributions of electron number density $n_e(x,z)$ and $n_e(y,z)$ in critical units $n_c$, see Tab.~\ref{tab:units}.
	
	\item Spatial distributions for all emerged ionization states of nitrogen  $n_{Nq}(y,z)$ ($q=0,+1,\dots,+7$) in critical units $n_c$.  
	
	\item Spatial distribution of electron kinetic energy $w_e(x,z)$ and $w_e(y,z)$ in units $n_cm_ec^2$.
	
	\item Dependence of the EM field intensity on space, such as $I(x,y=0,z)$, $I(x=0,y,z)$ - as was done in Section \ref{las_param} for vacuum.
\end{enumerate}

For item 1, we use electron momentum binning, setting only one bin (size of $\Delta p_{x,y}=0.1$ in $m_ec$ units) for $p_y=0\pm\Delta p_y/2$ or $p_x=0\pm\Delta p_x/2$ and as many bins as we need for the other two components to obtain probability densities $ f_1(p_x,p_z)$ or $f_2(p_y,p_z)$ respectively. Then we move on to  energy-angular variables
	\begin{align}
		\frac{dN^{(iz)}}{d\mathcal{E}_{K}d\theta}&=f_1(p_i,p_z)\sqrt{p_i^2+p_z^2+1},\; i=\{x,y\}
	\end{align}
where $p_{\{x,y,z\}}$ and $\mathcal{E}_K$ are in $m_ec$ and $m_ec^2$ units respectively, and in both planes, angle $\theta$ is measured from the direction $z$, see Fig.~\ref{hnap}. To make the grid of the obtained distributions uniform, we use 2D-interpolation. Also, we distinguish distributions for forward $(p_z>0)\cup(z>10\lambda_0)$ and backward $(p_z<0)\cup(z<-10\lambda_0)$ moving particles, thus, in the geometry of the numerical experiment $\theta\in(-73^\circ, 73^\circ)$.

Whereas the data related to the electron spectra can be measured experimentally and results of our simulations can be compared with them, the remaining items 2-5 on this list are needed rather to understand the acceleration mechanism and, therefore, to explain the features of these spectra. For the same purpose, we monitor other scalar parameters that evolve throughout the duration of the simulations: maximum absolute field values, integral energies of particles and fields, average charges and electron number densities.

\subsection{Simulation results and discussion}

We begin with a brief review of the experimental results reported in Ref. \cite{vallieres2022}. The experiment was performed at the Advanced Laser Light Source facility for IR ($\lambda_0=1.8$ $\mu$m) pulse of duration 12 fs and energy up to 2.8~mJ. Tight focusing was achieved using a HNAP with NA close to 1 (half-sphere focusing) and three independent radiation detectors measured the dose rate and its angular distribution. From processing the readings of these detectors, authors concluded that the radiation is characterized by high anisotropy in the forward direction with an estimated half-cone divergence angle $\theta_\mathrm{FWHM} \leq 17^\circ$, see Fig. ~\ref{hnap}, and a maximum kinetic energy of electrons $0.8\leq\mathcal{E}^{\mathrm{max}}_{K}\leq1.4$~MeV. In the geometry used in this article, these measurements refer to the $(y,z)$ plane.

For our preliminary PIC-simulation presented in the same paper \cite{vallieres2022}, the built-in paraxial beam was used, albeit with a narrow waist of 1~$\mu$m. In such modeling we observed predominantly forward accelerated electrons with maximum energies $\mathcal{E}_K^{\mathrm{max}}(a_0=4.86) \approx 2.6$~MeV and $\mathcal{E}_K^{\mathrm{max}}(a_0=3.08) \approx 1.9$~MeV. Also, although the angular distributions had a conical shape, their semicone divergence angle visually exceeded 40$^\circ$. Note that our simulations do not account for electron propagation in air to the detector, which may be located several meters away from the HNAP. Thus, one would naturally expect additional accumulation of angular divergence during the propagation process, which means that even lower values for this parameter than those recorded in the experiment would be expected in-situ. In the next two subsections, we not only demonstrate that using a model with a properly described tightly-focused pulse results in better agreement with experiment, but also perform a parameter scaling to better understand the acceleration mechanism.

\subsubsection{Scaling with $\lambda_0$ at $a_0=5$}\label{scl_a0}
\begin{figure}[t]
	\centering{
		\includegraphics[scale=1,trim={0 0 0 0},clip]{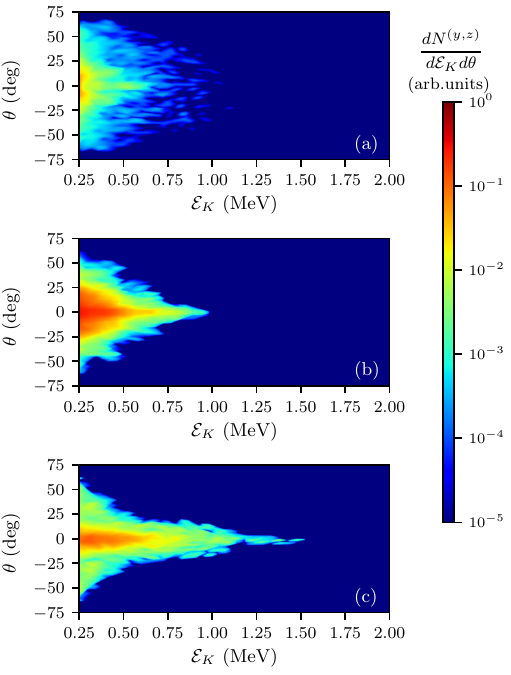}}
	\caption{Energy-angular spectra of electrons accelerated forward by a pulse with $a_0=5$ are shown at $t\approx172\cdot T_r$ for $\lambda_0$: (a) 0.8 $\mu$m ($t\approx73$ fs), (b) 1.8 $\mu$m ($t\approx165$ fs), (c) 2.5 $\mu$m ($t\approx229$ fs). }  \label{elec_spec2d}
\end{figure}

\begin{figure}[h]
	\centering{
		\includegraphics[scale=1,trim={0 0 0 0},clip]{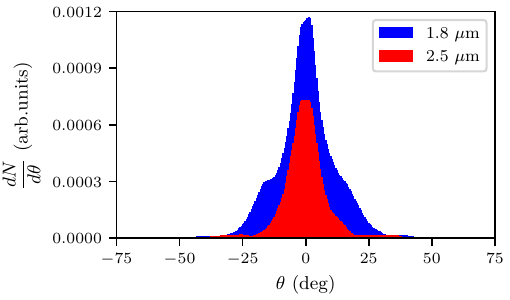}}
	\caption{Angular distributions of electrons for $\lambda_0 =1.8$ and 2.5 $\mu$m. Half-cone divergence angles $\theta_\mathrm{FWHM}$ are 13.8$^\circ$ and 12.6$^\circ$ respectively.}  \label{Half_cone}
\end{figure}

Figure~\ref{elec_spec2d} shows the results for the spectra in the $(y,z)$ plane of electrons accelerated by a tightly-focused laser pulse with $a_0=5$, at three different $\lambda_0$. First, in inset (b) we observe that the maximum kinetic energy of the electrons now falls within the experimentally determined interval. Moreover, in all plots (a)-(c) the maximum kinetic energies of electrons in the simulations are below the limiting value given by the formula for the ponderomotive force \eqref{max_kin}: $\mathcal{E}_K^{\mathrm{max} }(a_0 =5)\approx2.1$~MeV. 
In addition, the angular distribution is anisotropic: the measure of the angle of divergence $\theta_\mathrm{FWHM}$  does not exceed 14$^\circ$ for $\lambda_0=1.8$ $\mu$m and is even narrower for $\lambda_0=2.5$ $\mu$m, see Fig.~\ref{Half_cone}. 
Again, this is in agreement with the experimental results for $\lambda_0=1.8$ $\mu$m, thus increasing our confidence in the methodology and simulations results.
Lastly, for all $\lambda_0$ we observe predominantly forward-accelerated electrons. As for backward-moving particles with $\mathcal{E}_K>0.25$~MeV (minimum kinetic energy detectable in the experiment), they are unobservable at $\lambda_0\geq 4.5$ $\mu$m. For shorter wavelengths, only a small number of electrons randomly distributed in $\theta$ (with $\mathcal{E}_K^{\mathrm{max}}\leq 0.65$~MeV at $\lambda_0=0.8$ $\mu$m and $\mathcal{E}_K^{\mathrm{max}}\leq0.35$ MeV at $\lambda_0=1.8$, $2.5$ and $3.5$ $ \mu$m) is observed.

\begin{figure}[t]
	\centering{
		\includegraphics[scale=1,trim={0 0 0 0},clip]{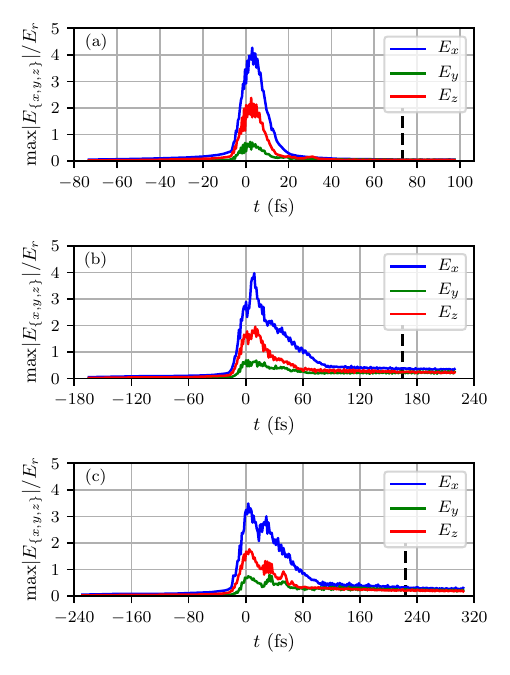}}
	\caption{Maximum over the entire simulation box values of the electric field components in time for pulse with $a_0=5$ and $\lambda_0$ (a) 0.8 $\mu$m, (b) 1.8 $\mu$m and (c) $2.5$ $\mu$m. Vertical dashed lines show the times at which the spectra were taken in Fig.~\ref{elec_spec2d}. }  \label{abs_max}
\end{figure}

The maximum energies below the theoretical value of 2.1 MeV for all $\lambda_0$ can be partly explained by decreased values of $E_x$ compared to the value $a_0=5$ in vacuum, as seen Fig.~ \ref {abs_max} which plots the maximum amplitude of all electric field components in the simulation box as a function of time. The most probable mechanisms explaining this behavior are ionization and reflection losses as the pulse propagates through the plasma. As for the $E_y$ and $E_z$-components in Fig.~\ref{abs_max}(a)-(c), their maximum values do not exceed the corresponding maximum values in vacuum. However, these components decrease more slowly in plasma, demonstrating interaction with electrons and their back-reaction on the electromagnetic field. 

From Figure~\ref{abs_max} we also see, that the chosen final time for Fig.~\ref{elec_spec2d} of $173\cdot T_r$ is suitable to acquire the spectra. By this time, the field amplitude is greatly reduced, the acceleration process is completed, and the particles are free-streaming towards the boundaries of the simulation box. For example, in case of $\lambda_0=2.5$ $\mu$m, see insert~(c), the electron spectrum does not change after $\approx190$ fs. Finally,  we obtained very similar spectra for the $(x,z)$ plane (not shown here for simplicity), perhaps with a slightly wider base at $\mathcal{E}_K<0.3$ eV, so we can conclude that electron emission has a conical shape.


\begin{figure*}
	\centering{
		\includegraphics[scale=1, trim={0 0 0 0},clip]{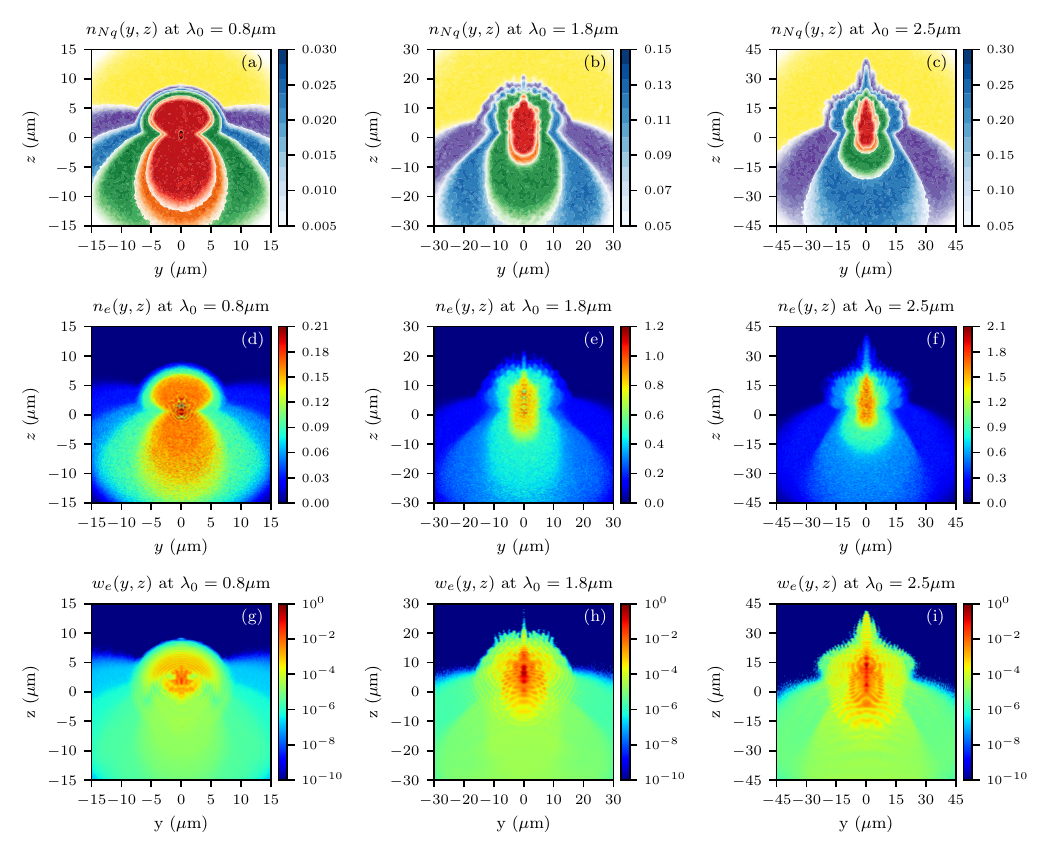}}
	\caption{Spatial distributions near acceleration region at time $t\approx57\cdot T_r$ for (a)-(c): nitrogen ions number densities, $n_{Nq}(y,z)$ ($n_c$ units), ions species are shown in colors: N$^0$ - yellow, N$^{1+}$ - purple, N$^{2+}$ - blue, N$^{3+}$ - green, N$^{4+}$ - orange, N$^{5+}$ - red, N$^{6+,7+}$ - brown, dark red (colorbars are shown for N$^{2+}$ only, assuming the same scales for all ion species for a given $\lambda_0$); (d)-(f) electrons  number density $n_e(y,z)$ ($n_c$ units) and (g)-(i) electron kinetic energy $w_e(y,z)$ ($n_cm_ec^2$ units).	The wavelengths are as follows: (a),(d),(g) - $\lambda_0=0.8$ $\mu$m ($t\approx24$ fs), (b),(e),(h) - $\lambda_0=1.8$ $\mu$m ($t\approx55$ fs) and (c),(f),(i) - $\lambda_0=2.5$ $\mu$m ($t\approx76$ fs).}  \label{all_dens}
\end{figure*}

\begin{figure}[h!]
	\centering{
		\includegraphics[scale=1,trim={0 0 0 0},clip]{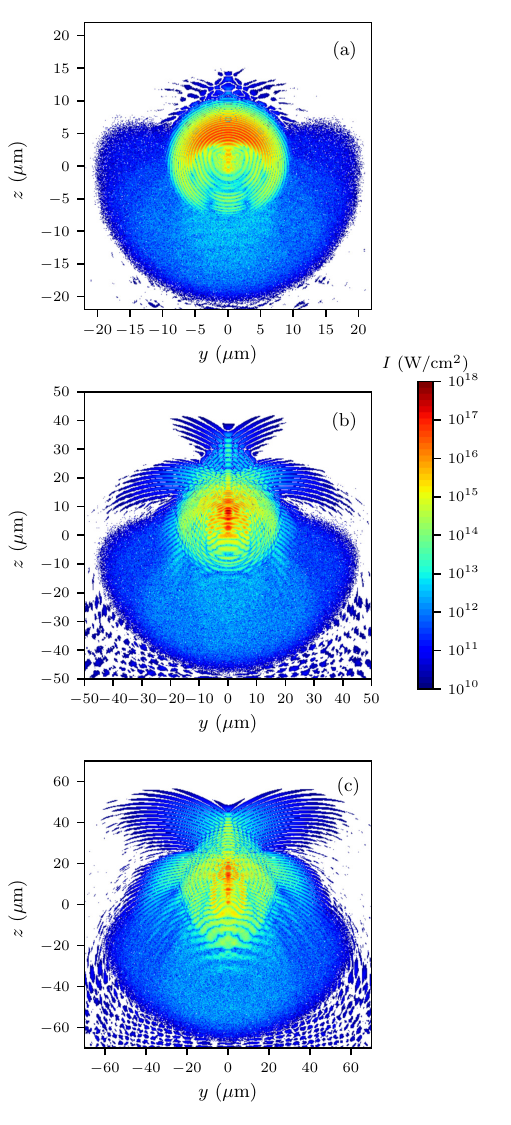}}
	\caption{Intensity $I(0,y,z,t)$ of the self-consistent field in plasma  at $t\approx57\cdot T_r$ for (a) $\lambda_0=0.8$ $\mu$m ($t\approx24$ fs), (b)~$\lambda_0=1.8$ $\mu$m ($t\approx55$ fs) and (c) $\lambda_0=2.5$ $\mu$m ($t\approx76$ fs).}  \label{intens_3plas}
\end{figure}

In our model, the electron acceleration efficiency depends on several factors: the degree of air ionization, the effective conversion from EM field energy into electron kinetic energy oscillating in this field, and the resulting action of the cycle-averaged relativistic ponderomotive force. Let us look at the details of the acceleration process, having data from the simulation of electron distributions - Fig.~\ref{all_dens} and fields - Fig.~\ref{intens_3plas}.


For a fixed value of $a_0$, as the wavelength decreases, the intensity increases in a region of scale $\lambda_0$ surrounding the focus, as seen in Fig.~\ref{focus_xy}. Hence, higher ionization states of air molecules can be reached and more electrons can potentially participate in the acceleration process. The energy conversion between the laser field and electrons crucially depends on the local electron number density: the closer $n_e$ is to the critical value $n_c$ at a given wavelength, see Tab~\ref{tab:units}, the more electrons can effectively interact with the wave. 
Because the ions are fixed, we can estimate the electron density at the moment where they are ionized from their parent ion by 
\begin{align}\label{elec_nc}
\nonumber  \frac{n_e(x,y,z)}{n_c} &\sim (n_{\mathrm{N}0} Z_\text{N}+n_{\mathrm{O}0} Z_\text{O})\cdot \frac{e^2\lambda_0^2}{4\pi^2c^2\varepsilon_0m_e} \\
&\sim n_{e1}(0.79 Z_\text{N}+0.21 Z_\text{O}),
\end{align}
where coefficient $n_{e1}=0.048[\lambda_0(\mu\text{m})]^2$, and $Z_\text{N}$ and $Z_\text{O}$ are the local (for brevity in their notation we omitted $(x,y,z)$) degrees of ionization for nitrogen (N$^{+Z_\mathrm{N}}$) and oxygen (O$^{+Z_\mathrm{O}}$), see Fig.~\ref{all_dens}(a)-(c). For example, in the case of single local ionization of nitrogen and oxygen, we have $n_e(x,y,z)/n_c=n_{e1}$. For convenience, Table~\ref{tab:dens} contains the coefficients $n_{e1}$ for different wavelengths. 
In Figure~\ref{all_dens}(a)-(c), for all $\lambda_0$ we observe at least ions N$^{5+}$, O$^{5+}$ and even higher ionization states in the focal region. Therefore, using Eq. \eqref{elec_nc} and the numbers in Table \ref{tab:dens}, we expect regions of near-critical and even over-dense plasma for $\lambda_0\geq 1.8$~$\mu$m. This is confirmed by looking at the actual electron density obtained in the simulation in Figs.~\ref{all_dens} (d)-(f).
We can now look at how this affects the acceleration processes.

\begin{table}[h]
	\caption{Electron number density $n_{e1}$ in critical units in the case of single ionization, see formula \eqref{elec_nc}.  \label{tab:dens}}
	\begin{ruledtabular}
		\begin{tabular}{ l c c c c c c c c} 
			$\lambda_0$ ($\mu$m) & 0.8 & 1.8& 2.5 & 3.5&4.5&5.5&7.0\\	
			\hline
			$n_{e1}$ ($n_c$) & $0.03$ & 0.16 & 0.30 & 0.59 & 0.98 & 1.46 &2.36
		\end{tabular}		
	\end{ruledtabular}
\end{table}

\begin{figure}[h]
	\centering{
		\includegraphics[scale=1,trim={0 5 0 5},clip]{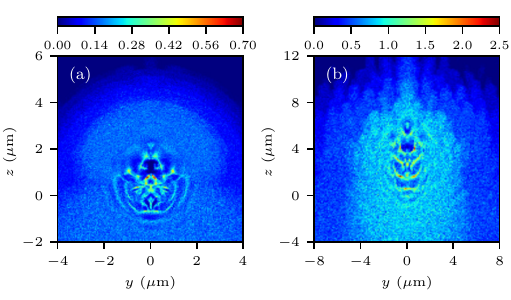}}
	\caption{Electron number densities $n_e(y,z)$ ($n_c$ units) near the laser beam focus at $t=28\cdot T_r$ for (a) $\lambda_0=0.8$ $\mu$m (12 fs) and (b) $\lambda_0=1.8$ $\mu$m (27~fs).}  \label{two_dens}
\end{figure}

For $\lambda_0=0.8$~$\mu$m, the density is far from the critical regime: $n_{e}^{\mathrm{max}} \approx 0.21$. Because of this low effective density, we can even notice the preserved structure of the laser pulse in Fig.~\ref{intens_3plas} (a), where the particles oscillate with $E_x $-component as seen in Fig.~\ref{all_dens} (g). However, if we look at the focal region with better resolution, as in Fig.~\ref{two_dens}(a),(b), we can observe the formation of a trail of microbubbles (each of size $\leq$ 1 $\mu$m) in the wake left after the laser pulse.
This separation of charges in space, which begins with one bubble at the moment of focusing, occurs due to the displacement of electrons by the laser ponderomotive force and leads to the appearance of a  static field component there, akin to the laser wakefield acceleration mechanism. As the data shows, the kinetic energies of particles accelerated in the region behind the laser pulse displayed in Fig.~\ref{all_dens}(g) are acquired predominantly due to the bubble's fields, since this area coincides with the position of microbubbles where oscillating $E_x$-component is low. However, the generated static field in this region is only 0.2$E_r$ and thus, is not responsible for accelerating electrons to the MeV range. 

Although neither the vacuum analytical solution \eqref{ExEyEzBxByBz} nor the numerical solution of the Strattocalculator for the LP pulse \cite{Vallieres:23} has an $E_z$ component in the entire $(y,z)$ plane, this component appears in the plasma.  On the optical axis, we actually observe oscillating electric fields (for all components) existing for long times after focusing for all $\lambda_0\geq1.8$ $\mu$m, but their amplitudes do not exceed 0.5$E_r$. Again this is not enough to accelerate particles to the MeV energy scale. Having ruled out these possible acceleration mechanisms, we proceed by exploring another one - direct acceleration by the relativistic ponderomotive force, which has already been proposed for these experimental conditions in \cite{vallieres2022}.

In the nonrelativistic limit, the ponderomotive force  
${\bf F}_p=-\dfrac{m_ec^2}{4E_r^2}\nabla{|E({\bf r})|^2}$,
accelerates charged particles in the direction of the electric field gradient. Then, in an underdense plasma, which does not strongly modify the field, e.g. when $\lambda_0=0.8$ $\mu$m see \ref{intens_3plas}(a), it can be safely assumed that electrons oscillate in the $E_x$ direction and slowly drift along the entire leading edge of the pulse, thus yielding a smooth electron spectra with a maximum at $ \theta=0$ in the forward direction, and a similar spectra along the trailing edge for backward particles. In the relativistic regime, expressions for the ponderomotive force are more sophisticated and non-universal, so, the behavior of particles is more complex, with the possibility of falling into chaos \cite{Bauer, Bituk, Lau}. In the relativistic regime, oscillating particles acquire a $p_z$ component not so much due to the gradient part of the ponderomotive force, but due to its other part, the magnetic term $e({\bf v}_x/c)\times{ \bf B}$, which makes particles move forward in the $z$ direction rather than backward, see e.g. \cite{Pukhov}. Thus, we have already given one argument in favor of direct acceleration by the ponderomotive force when discussing the spectra in Fig.~\ref{elec_spec2d}, by noting that electrons were accelerated predominantly in the forward direction with low divergence angle. In addition, as seen in Figs.~\ref{all_dens}-(g) and \ref{intens_3plas}-(a), the electrons are initially accelerated along the wavefront following the field gradient as predicted by the ponderomotive force. Once they reach a high enough velocity, the magnetic term starts to be important and push the electrons in the forward direction.    

In contrast to the case with $\lambda_0=0.8$ $\mu$m, for $\lambda_0=1.8$ and 2.5 $\mu$m, the leading edge of the pulse is strongly reflected when colliding with a region of over-dense plasma, see \ref{intens_3plas}(b),(c).  Due to the spatial redistribution of wave intensity, we observe a strong distorsion of the spherical wave front further from the optical axis, whereas along the $z$-axis the strength of the self-consistent field increases significantly compared to propagation in vacuum, e.g., at $\lambda_0=1.8$ $\mu$m at time $57\cdot T_r$ it is $5.5\cdot10^{17}$ W/cm$^2$ versus $1.3\cdot10^{16}$ W/cm$^2$. This effect enhances electron acceleration because the particles are subjected to higher fields and gradients.  

\begin{figure}[t]
	\centering{
		\includegraphics[scale=1,trim={0 10 0 10},clip]{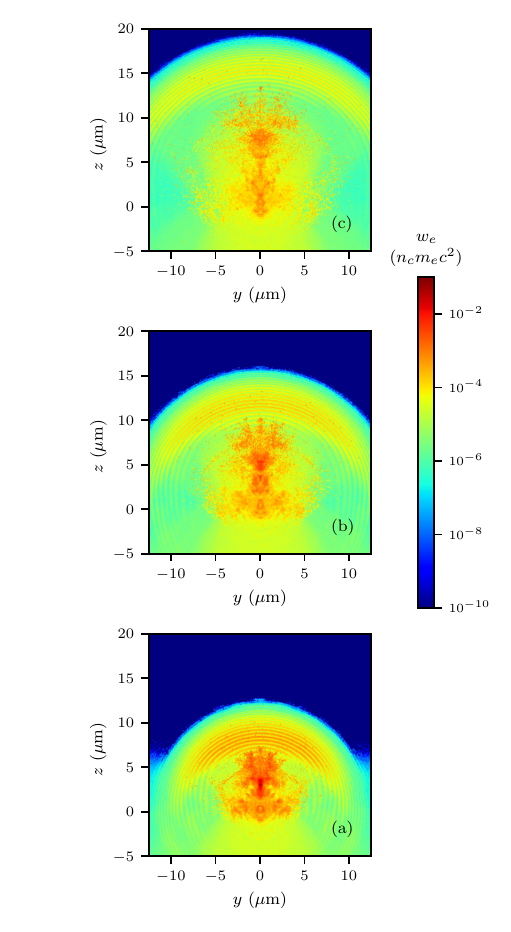}}
	\caption{Distributions of electrons kinetic energy $w_e(y,z)$ in the field of a tightly-focused laser pulse with $\lambda_0=0.8$ $\mu$m at times (a) $t=37$ fs, (b) $t=49$ fs and (c) $ t=61$ fs.}  \label{three_kin1}
\end{figure}

\begin{figure}[t]
	\centering{
		\includegraphics[scale=1,trim={0 10 0 10},clip]{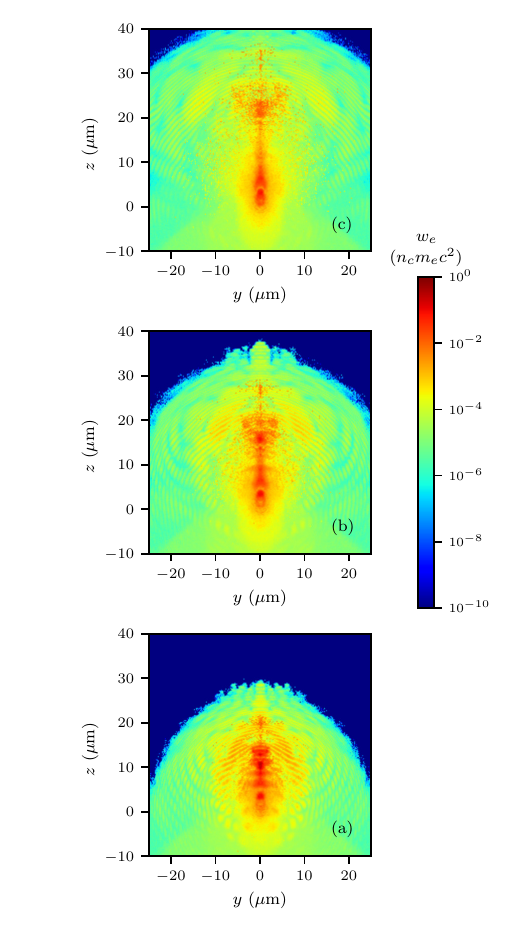}}
	\caption{Distributions of electrons kinetic energy $w_e(y,z)$ in the field of a tightly-focused laser pulse with $\lambda_0=1.8$ $\mu$m at times (a) $t=82$ fs, (b) $t=110$ fs and (c) $ t=137$ fs.}  \label{three_kin2}
\end{figure}

In Figures ~\ref{three_kin1},\ref{three_kin2}(a)-(c) we can trace the evolution of the kinetic electron density shown in Fig.~\ref{all_dens}(g),(h). The beams of electrons moving forward are clearly visible, and as a result we obtain the broader in angles spectra with $\theta_\mathrm{FWHM}\approx22^\circ$ for $\lambda_0=0.8$ $\mu$m (accelerated along the entire undeformed wave front), see Fig.~\ref{elec_spec2d}(a), and sharply conical angular electron spectra for $\lambda_0\geq1.8$ $\mu$m, see Fig.~\ref{elec_spec2d}(b),(c) and Fig.~\ref{Half_cone}. In a plasma with a density close to critical, more electrons oscillate with the $E_x$ component and effectively acquire $p_z$ due to the relativistic ponderomotive force, so we see more particles in the spectra at $\lambda_0=1.8$ and 2.5 $\mu$m, than for $\lambda_0=0.8$~$\mu$m.
Figure~\ref{three_kin2} also shows that in addition to the beam of electrons accelerated forward, some of them remained near the focus, oscillating and generating currents there (see in Fig.~\ref{abs_max}(b))  with an amplitude of $\approx 0.4E_r$, which is preserved up to $ t>120$ fs. These laser-excited oscillations arise in a region where there is an over-dense plasma and cannot propagate, since their group velocity is $v_g=\mathrm{Re}\{c\sqrt{1-\omega_{pe}^2/\omega_0^2}\}=\mathrm{Re}\{c\sqrt{1 -n_e/n_c}\} =0$, whereas electrons moving forward and oscillating with the laser pulse travel on the edge of the region between over-dense and under-dense plasma.

\begin{figure}[t]
	\centering{
		\includegraphics[scale=1,trim={0 0 0 0},clip]{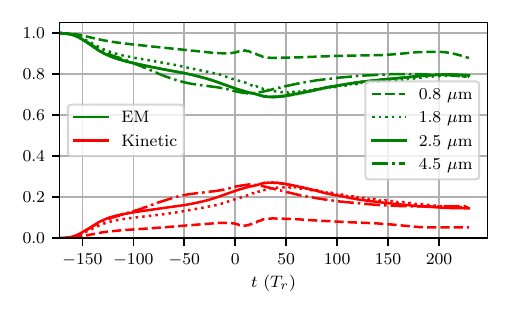}}
	\caption{Normalized EM and kinetic energies integrated over the simulation box in time for  $\lambda_0=\{0.8, 1.8, 2.5. 4.5\}$ $\mu$m.}  \label{Energy}
\end{figure}

In terms of energy conversion, approximately 30-35$\%$ of EM energy integrated over the simulation box, is spent to air ionization and electron acceleration when $\lambda_0\geq1.8$ $\mu$m, but it is only slightly above 10$\%$ in the case of $\lambda_0=0.8$ $\mu$m, see Fig.~\ref{Energy}. As we have just seen, only part of this energy turns into the energy of a directed electron beam for all $\lambda_0$. When $\lambda_0\geq 1.8$ $\mu$m, i.e. when the transfer of EM energy to electron acceleration is sufficiently effective, we observe a shift in the position of the maximum kinetic energy of electrons (or minimum EM energy) to the left with increasing wavelength. The larger $\lambda_0$, the more efficiently EM energy is spent accelerating electrons before focusing. However, at large $\lambda_0$, the acceleration process loses its effectiveness due to an increase in the size of the overdense plasma region, where the pulse can no longer propagate and accelerate electrons. 

\begin{figure}[t]
	\centering{
		\includegraphics[scale=1,trim={0 0 0 0},clip]{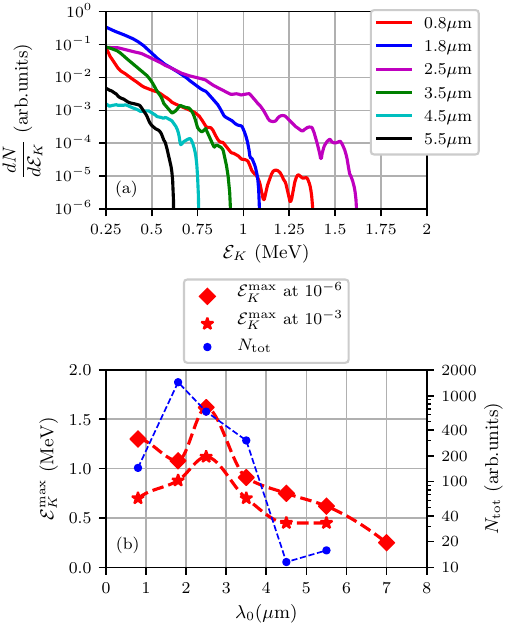}}
	\caption{(a) Energy spectra of electrons accelerated by a pulse with $a_0=5$ and averaged over $\theta$ in the $(x,z)$ and $(y,z)$ planes are shown at $t\approx73$ fs ($\lambda_0=0.8$ $\mu$m), $t\approx165$ fs ($\lambda_0=1.8$ $\mu$m),  $t\approx229$ fs ($\lambda_0=2.5$ $\mu$m), $t\approx320$ fs  ($\lambda_0=3.5$ $\mu$m) and $t\approx480$ fs  ($\lambda_0=4.5$ $\mu$m). (b) The dependencies of $\mathcal{E}_K^{\mathrm{max}}$ on $\lambda_0$ are shown in red from the data presented in inset (a) for two levels in $dN/d\mathcal{E}_K$. Blue circles show the dependence of the electron yield $N_\mathrm{tot}$ on $\lambda_0$ with a cutoff at 250 keV, the values are shown on the additional $y$-axis to the right.
	}  \label{forw_all}
\end{figure}

Finally, we summarize the results of Section~\ref{scl_a0} in Fig.~\ref{forw_all}. In insert (a) we carried out an averaging over the $\theta$-angle in both planes $(x,z)$ and $(y,z)$ to obtain the particle spectra. In the case of the $(x,z)$ plane, the maximum energy is usually higher by about 5~$\%$, probably because in this plane we have, in addition to the $E_x$ component, a comparable $E_z$ component in vacuum. Here, it should be noted that the number of superparticles representing electrons are different in each spectrum, since it depends on the peak intensity of the ionizing beam, which decreases with wavelength. Thus, for $\lambda_0=0.8$ $\mu$m we have created $1.3\cdot10^{10}$ superparticles, for $\lambda_0=1.8$ $\mu$m - $5.8\cdot10^{9}$, and for $\lambda_0=5.5$ $\mu$m there is only $10^{9}$ of them. This can cause some statistical fluctuations, which probably explain the humps in the tail of the spectrum at $\lambda=0.8$ $\mu$m. 

In Fig.~\ref{forw_all}~(b) the simulated maximum electron kinetic energies are plotted as functions of $\lambda_0$ at two spectral levels $dN/d\mathcal{E}_{K}$, thus modelling different detector sensitivities. 
For both levels we can observe an increase in $\mathcal{E}_K^\mathrm{max}$ for $\lambda_0\in[1.8, 2.5]$ $\mu$m, followed by a decline for $\lambda_0 > 2.5$ $\mu$m. Indeed, the peak intensity of the beam decreases with increasing wavelength, which leads to a decrease in the number of ionization electrons. Moreover, not all ionization electrons can be accelerated efficiently; the parameter responsible for the conversion of EM energy into kinetic energy of electrons, equal to the ratio $n_e/ n_c\propto n_e\lambda_0^2$, increases with wavelength of the laser beam. As a compromise between these opposing trends, for a fixed $a_0=5$ we observe a maximum of $\mathcal{E}_K^\mathrm{max}$ achieved at $\lambda_0=2.5$ $\mu$m. 

Insert (b) also shows the total number (yield) of accelerated electrons calculated from the data in inset (a). For high energy electrons with a cutoff at 250 keV as in inset (a), the ratio between the yield at $\lambda_0=1.8$ $ \mu$m and at $\lambda_0=0.8$ $\mu$m reaches a value of $N(\lambda_{0} = 1.8\; \mu\mbox{m})/N(\lambda_{0} = 0.8\; \mu\mbox{m}) \approx 10$. Therefore, although it is possible to reach similar kinetic energies at $\lambda_0=1.8$ $\mu$m and $\lambda_0=0.8$ $\mu$m, the yield is much lower at $\lambda_0=0.8$ $\mu$m.
Furthermore, the yield decreases above $\lambda_0=1.8$ $\mu$m, so that at $\lambda_0=2.5$ $\mu$m it is 2.2 times less than the maximum - primarily due to the energy region below 0.6 MeV. As follows from Tab.~\ref{tab:dens}, for $\lambda_0>1.8$ $\mu$m the focusing region is filled with over-dense plasma if the ionization levels of N and O reach 3-4. Thus, for longer wavelengths, not only do we have few ionizing electrons, but our ability to effectively accelerate them with a tightly-focused beam is reduced.

\subsubsection{Scaling with $a_0$ at $\lambda_0=2.5$ $\mu$m}
\begin{figure}[t]
	\centering{
		\includegraphics[scale=1,trim={0 0 0 0},clip]{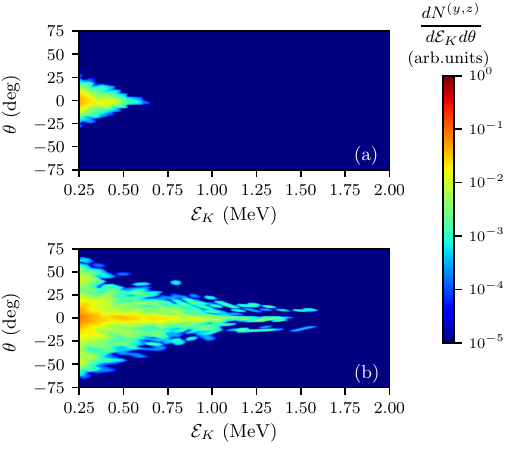}}
	\caption{Energy-angular spectra of electrons accelerated forward by pulses with $\lambda_0 = 2.5$ $\mu$m are shown at $t\approx229$ fs for field amplitudes (a) $a_0=3.6$ and (b) $a_0=7$.}  \label{elec_spec2d_25}
\end{figure}

\begin{figure}
	\centering{
		\includegraphics[scale=1,trim={0 0 0 0},clip]{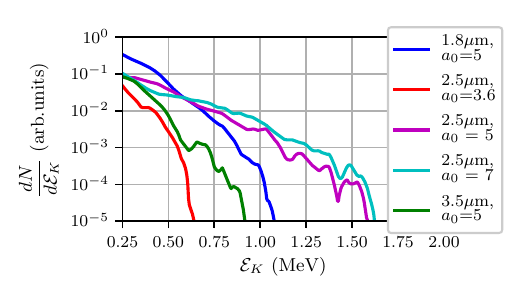}}
	\caption{Energy spectra of electrons accelerated by laser pulses with $\lambda_0=2.5$ $\mu$m and different field amplitudes: $a_0=3.5$, $a_0=5$ and $a_0=7$ averaged over $\theta$ in the $(x,z)$ and $(y,z)$ planes are presented at $t=229$ fs. To compare, we show the spectra with parameters $\lambda_0=1.8$ $\mu$m, $a_0=5$ and $\lambda_0=3.5$ $\mu$m, $a_0=5$ related to times $t\approx229$ fs and  $t\approx320$ fs, respectively.}
 	\label{forw_25}
\end{figure}

Since we obtained the maximum kinetic energy and a large number of accelerated electrons for $\lambda_0=2.5$ $\mu$m, we continue to use this wavelength to simulate the spectra at lower and higher $a_0$, see fig.~\ref{elec_spec2d_25}(a),(b). From these two plots and Fig.~\ref{forw_25} we see a noticeable decrease in $\mathcal{E}_K^{\mathrm{max}}$ when we decrease $a_0$ from 5 by $\approx1.4$. 
On the other hand, we observe only a small gain in $\mathcal{E}_K^{\mathrm{max}}$, if we increase $a_0$ up to value of 7. By setting $a_0=7$, we get almost the same $I_0$ as when $\lambda_0=1.8$ $\mu$m and $a_0=5$. For $a_0=3.6$, $I_0$ is close to the one for $\lambda_0=3.5$ $\mu$m with $a_0=5$. So this time for $a_0=3.6$ the $n_e/n_c$ ratio decreases due to decreased ionization, so the energy conversion is less efficient for the same $\lambda_0=2.5$ $\mu$m, making the acceleration even less efficient than if just the value of $a_0$ was decreased. 
On the contrary, at $a_0=7$ we have much higher ionization compared to $a_0=5$, which leads to larger regions of over-dense plasma and hence stronger wave reflection. These effects limit the benefit of having higher intensities by increasing $a_0 $.

\section{Conclusion}\label{concl}
We numerically simulated the acceleration of electrons in tightly-focused ultrashort IR laser pulses in ambient air. The approach is based on PIC simulations with the use of an exact vacuum analytical solution in combination with a spherical gas target. Our modelling confirms the hypothesis that the ponderomotive force plays a fundamental role in the acceleration of electrons under these experimental conditions \cite{vallieres2022}. 

The efficiency of the kinetic energy gain depends on the normalized field amplitude $a_0$ and the electron number density in critical units $n_e/n_c$. Since at given $\lambda_0$, $a_0\propto\sqrt{I_0}\lambda_0$ and the number density of ionization electron increases with $I_0$, whereas $n_c\propto\lambda_0^{-2}$, both factors $a_0$ and $\sqrt{I_0 }\lambda_0$ work to increase $\mathcal{E}_K^{\mathrm{max}}$. However, when the region of over-dense plasma becomes large due to ionization, the reflected laser pulse cannot propagate in this region and hence, the electron kinetic energy stops increasing with $a_0$. For this reason, for $\lambda_0=2.5$ $\mu$m we see almost the same $\mathcal{E}_K^{\mathrm{max}}$ as at $a_0=5$ and $a_0=7$.
Similarly, if we fix the value $a_0$, the maximum electron energy increases with $\lambda_0$. Owing to reflections of the pulse on the over-dense plasma at larger $\lambda_{0}$, the kinetic energy starts to decrease.
Thus, in ambient air simulations at $a_0=5$, an optimal value for $\mathcal{E}_K$ was detected at $\lambda=2.5$ $\mu$m. 
According to our analysis, to reach higher energies which require larger field amplitudes $a_0$, the air pressure should be decreased. For example, in the case of $\lambda_0=5$ $\mu$m and intensity $I_0\approx 0.54\cdot 10^{19}$ W/cm$^2$, which leads to $a_0=10$, at pressure $n_{e1}(5\mu\mathrm{m})/n_{e1}(2.5\mu\mathrm{m})\approx4$ times less than under standard conditions, the maximum electron energy in our simulations reaches 3 MeV. In the general case,  at given $a_0$, the wavelength at which maximum kinetic energy of electrons can be reached, depends on the gas type and possible ionization levels. Thus, for other gases, the scaling law we see here could be different. This will be studied in other articles.
Also, for future research, we aim to combine PIC codes with the numerical solution of the Strattocalculator \cite{Dumont_2017}, and considering electron acceleration in tightly focused radially polarized (RP) pulse.

\section*{Acknowledgments}
Authors thank Shanny Pelchat-Voyer for sharing her unpublished results and codes to compute the exact April's solution for radially polarized pulse in vacuum.
This research was enabled in part by support provided by the Digital Research Alliance of Canada  \url{alliancecan.ca.}

\bibliography{mybibfile}

\end{document}